%% file: CahoyMainFINAL.tex
\documentclass[doublespacing, 12pt, authoryear]{article}

\usepackage[latin1]{inputenc}
\usepackage{graphicx, amsmath, amsthm,amsfonts, amssymb, subfigure, amstext}
\usepackage{wrapfig, amsopn, latexsym,  epsfig, multirow,  verbatim}
\usepackage{comment,  caption,  bm}

\title{Estimation of Mittag-Leffler parameters}
\author{Dexter O. Cahoy\\
Department of Mathematics and Statistics\\
College of Engineering and Science\\
Louisiana Tech University \\
Ruston, LA 71271\\
\texttt{dcahoy@latech.edu}}

\date{}

\allowdisplaybreaks

\usepackage[colorlinks, citecolor=red,linkcolor=blue]{hyperref}

\begin{document}

\maketitle

\begin{abstract}
\indent

We propose a procedure for estimating the parameters of the Mittag-Leffler (ML) and the generalized Mittag-Leffler (GML) distributions. The algorithm  is less restrictive, computationally simple and necessary to make these models usable in practice.  A comparison with the fractional moment estimator indicated favorable results for the proposed method.

\vspace{0.1in}

\end{abstract}

\input{s1b.tex}
\input{s2b.tex}

\input{s3b.tex}

\section{Acknowledgment}
The author is grateful to the Associate Editor and the reviewer for their insightful comments and suggestions that led to the improvement of the article. This research was supported under the Louisiana Board of Regents Research Competitiveness Subprogram grant LEQSF(2011-14)-RD-A-15.

\end{document}

%% file: s1b.tex
\section{Introduction} \label{a}

For the last  two decades, the Mittag-Leffler  function has gained popularity in many scientific areas. For instance, the Mittag-Leffler (ML) distribution  originally introduced by \cite{pil90} has now been used to model  random phenomena in  finance and economics \cite{mas06, sca06}. In addition, \cite{koz01} studied  the  Mittag-Leffler  ML$(\alpha, \delta)$  distribution where  the integral and series representations of the probability density function  are
\[
i.) \quad f_T (t) =  \frac{1}{t} \int\limits_0^\infty e^{-\xi} g\left[ t /( \delta^\alpha \xi) \right]d\xi   \qquad \text{and} \qquad  ii.) \quad  f_T (t) =   \frac{ t^{\alpha-1}}{\delta^\alpha} E_{\alpha, \alpha} \left[-(t/\delta)^\alpha\right],
\]
correspondingly, $\delta>0$ is the scale parameter, $0 < \alpha \leq 1,  g(\eta) =\sin (\alpha \pi) / [ \pi( \eta^\alpha + \eta^{-\alpha} + 2\cos (\alpha \pi ) ],$ and
\[
 E_{\alpha, \nu} (\tau)=\sum\limits_{k=0}^\infty \frac{\tau^k}{\Gamma (\nu+ k \alpha)}
\]
is the generalized Mittag-Leffler function. The ML$(\alpha, \delta)$ density function has  the Laplace transform
\[
 \widetilde{f}_{\alpha, \delta}(\lambda) = \int_0^\infty e^{-\lambda t} f_T (t) dt= \left[1 + (\delta \lambda)^\alpha \right]^{-1},
\]
and is showed to be geometric stable.  It can also be easily seen as a generalization of the density of an exponential distribution with parameter $\delta > 0$. Furthermore, \cite{koz01} constructed the following structural representation of a ML$(\alpha, \delta)$  distributed random variable $T$ as
\begin{equation*}
T\stackrel{d}{=} \delta Z R^{1/\alpha},  \tag{1.1}    \label{1.1}
\end{equation*}
where $Z$ has the standard exponential distribution $\exp (1)$, and $R$ has the density function
\[
f_R(r)= \frac{\sin (\pi \alpha)}{\alpha \pi \left[ r^2 +2r\cos(\alpha \pi) +1 \right]}, \quad 0< \alpha <1, \; r >0.
\] 
It is also shown  that the  $qth$ fractional moment of the random variable $T$  is:
\begin{equation*}
\mathbf{E} T^q= \frac{q \pi \delta^q  }{\alpha \Gamma (1-q) \sin (\pi q/\alpha)}, \quad 0 <q <\alpha. \tag{1.2}    \label{1.2}
\end{equation*}
The parameter estimation problem for this model was first addressed by \cite{koz01}. They proposed  fractional moment estimators for the ML$(\alpha, \delta)$ distribution which require appropriate constants prior to the calculation of  the estimates.  Note that the pre-selection of appropriate values requires information about the true or unknown parameter $\alpha$ a priori, which is not feasible in practice. As
a direct consequence, it is expected that the above estimators will perform poorly when the restrictions are violated. 
Thus, it is mainly these drawbacks that motivate us to propose an estimation procedure  that avoids this difficulty and uses all the available data possible.

More recently, \cite{juslh10} used the generalized Mittag-Leffler (GML) distribution in astrophysics and  time series modeling. They specifically constructed GML  processes which are autoregressive time series models with GML as the stationary marginal distribution. Moreover, the cumulative distribution function of the generalized Mittag-Leffler distribution GML$(\alpha, \beta)$ is given by
\[
F_{\alpha, \beta} (x) = P[ X \leq x ] =\sum\limits_{k=0}^\infty \frac{(-1)^k \Gamma (k + \beta) x^{\alpha(k + \beta)}}{\Gamma (\beta) k! \Gamma (1 + \alpha(k + \beta ) )}, \qquad x>0.
\]
When $\alpha=1$, we get the gamma distribution while $\beta=1$ yields \cite{pil90}'s ML$(\alpha, 1)$ distribution. If $\alpha=\beta=1$, we get the exponential density. The GML probability density function has the specific form
\[
f_{\alpha,\beta} (x) = \int_0^\infty \frac{\alpha x^{\alpha \beta-1} e^{-(x/s)^\alpha} }{\Gamma ( \beta) s^{\alpha \beta}} d F_{ S_\alpha} (s),
\]
where $F_{ S_\alpha} (s)$ is the cumulative distribution function  of a strictly positive stable random variable  $S_\alpha$ with $\exp (- \lambda^\alpha)$ as the Laplace transform  of the corresponding probability density function. The  Laplace transform of the GML probability density function above is  $\widetilde {f}_{\alpha, \beta}(\lambda) = (1 + \lambda^\alpha)^{-\beta}.$ This distribution can be considered as the positive counterpart of \cite{pake98}'s generalized Linnik distribution with the probability density function having the Laplace transform $(1+|\lambda|^\alpha)^{-\beta},\; 0 < \alpha \leq 2, \; \beta >0$.  Furthermore, the mixture representation of a GML$(\alpha, \beta)$ distributed random variable $X$ is
\begin{equation*}
X \stackrel{d}{=}  W^{1/\alpha} S_\alpha,  \tag{1.3}    \label{1.3}
\end{equation*}
where $W$   is gamma distributed with scale parameter 1 and shape parameter $\beta$, and its probability density function given by
\[
f_W(w)=\frac{1}{\Gamma (\beta)}w^{\beta-1} e^{-w}.
\]
The fractional moments of the GML distributed random variable $X$  for $\beta=1$ are derived by \cite{pil90} as
\[
\mathbf{E}X^q= \frac{ \Gamma (1-q/\alpha) \Gamma (1 + q / \alpha) }{\Gamma (1-q)},
\]
while \cite{lin98a,lin98b} obtained the expression
\begin{equation*}
\mathbf{E}X^q= \frac{\Gamma (1-q/\alpha) \Gamma (\beta + q / \alpha) }{ \Gamma (1-q) \Gamma (\beta)}, \quad -\alpha \beta <q< \alpha, \tag{1.4}    \label{1.4}
\end{equation*}
for $0<\alpha \leq 1$ and $\beta >0$. Note that the moments are infinite for order $q \geq \alpha$. 

The main goal of this paper is to propose a procedure to estimate the model parameters in the  Mittag-Leffler (ML) and the generalized Mittag-Leffler (GML) distributions that uses all available information. This is necessary  in order for these models to be usable in practice. The rest of the paper is organized  as follows: In Sections 2 and 3, we  derive the first few moments of the log-transformed  Mittag-Leffler distributed random variables. In Section  4, we propose procedures  to estimate the parameters of  the ML$(\alpha, \delta)$ and the  GML$(\alpha, \beta)$ distributions.  In Section 5, some key points and extensions of our methodology are discussed. In Section 6,  some computational test results are shown and interval estimators  for the ML$(\alpha, \delta)$ parameters are derived using the  asymptotic normality of the point estimators.

%% file: s2b.tex
\section{Moments of the log-transformed ML random variable  $T$}

We now derive the first four log-moments of the random variable $T$.  Applying the log-transformation to the mixture representation (\ref{1.1}), we obtain
\begin{equation*}
T^{'} \stackrel{d}{=} \log (\delta ) + Z^{'} + \frac{1}{\alpha} R^{'},  \tag{2.1} \label{2.1}
\end{equation*}
where $T^{'}= \log (T)$, $Z^{'}=\log (Z)$, and $R^{'}=\log (R)$. 
Following \cite{cuw10}, it is straightforward to show the following first four non-central moments of  the random variables $Z^{'}$ and $R^{'}$:
\[
\mathbf{E}(Z^{'})=-\gamma, \qquad \mathbf{E}
\left( Z^{'} \right)^2=\gamma^2+\frac{\pi^2}{6},
\]
\[
\mathbf{E}\left(  Z^{'} \right)^3 = -
\gamma^3-\frac{\gamma\pi^2}{2}-2\zeta (3), \qquad
\mathbf{E}\left( Z^{'}\right)^4 =
\gamma^2\left(\gamma^2+\pi^2\right) +
\frac{3\pi^4}{20}+8\gamma\zeta (3),
\]
\[
\mathbf{E}\left(  R^{'} \right) = 0, \qquad  \mathbf{E} \left( R^{'}\right)^2=\frac{\pi^2}{3} \left(1- \alpha^2 \right),
\]
\[
\mathbf{E} \left(R^{'}\right) ^3=0, \quad \text{and} \qquad  \mathbf{E} \left(R^{'}\right)^4 = \frac{\pi^4}{15}\left( 7-10 \alpha^2+3 \alpha^4 \right).
\]
where  $\gamma=0.5772156649015328606065$ is the Euler's constant, and $\zeta (\tau)$ is the Riemann zeta function evaluated at $\tau$,

Taking the expectation of  (\ref{2.1}) and using the above moments, we get the mean and variance
\begin{equation*}
\mu_{T^{'}}= \log (\delta ) -\gamma, \quad \text{and} \quad \sigma_{T^{'}}^2= \frac{\pi^2}{6} \left( \frac{2}{\alpha^2} -1 \right), \tag{2.2} \label{2.2}
\end{equation*}
respectively. Observe that the mean $\mu_{T^{'}}$ does not involve the parameter $\alpha$ which is surprising and is due to the expected value of $R^{'}$ being zero.  Moreover, a similar calculation gives the third and fourth central moments as
\[
\mu_3^{'}=\mathbf{E} \left( T^{'} - \mu_{T^{'}}\right)^3= -2\zeta (3),\; \text{and} \;
\mu_4^{'}=\mathbf{E} \left( T^{'} - \mu_{T^{'}}\right)^4= \frac{\pi^4(\alpha^4-20 \alpha^2 +28)}{60\alpha^4},
\]
respectively which will be used in the  derivation of the interval estimates in the appendix.

\section{Moments of the log-transformed GML random variable $X$ }

Taking the logarithm of the mixture representation of the GML distributed random variable $X$ in (\ref{1.3}) yields
\[
X^{'} \stackrel{d}{=} \frac{1}{\alpha}W^{'} + S_\alpha^{'},
\]
where $X^{'} = \log (X),   W^{'}= \log (W)$,  $S_\alpha^{'}=\log (S_\alpha)$, and $S_\alpha$  is a one-sided  $\alpha$-stable distributed random variable with the Laplace transform of the probability density function  given as $\exp (-\lambda^\alpha), 0< \alpha \leq 1, \beta > 0$. From \cite{zol86}, the first four log-moments of $S_\alpha$ can be deduced as
\[
\mathbf{E}\left(  S_\alpha^{'} \right) = \mathbb{C}\left( \frac{1}{\alpha}-1\right), \qquad  \mathbf{E} \left( S_\alpha^{'}\right) ^2=\left(\frac{1}{\alpha}-1\right)^2 \mathbb{C}^2 +
\frac{\pi^2}{6}\left( \frac{1}{\alpha^2}-1\right),
\]
\[
\mathbf{E} \left(S_\alpha^{'}\right) ^3=\frac{-2(\alpha-1)^3\mathbb{C}^3+\mathbb{C}\pi^2
(\alpha-1)^2(1+\alpha)-4(\alpha^3-1)\zeta(3)}{2\alpha^3}, \quad \text{and}
\]
\[
\mathbf{E} \left(S_\alpha^{'}\right)^4 = \frac{1}{60}\bigg[
\bigg(\frac{1}{\alpha^3}-\frac{1}{\alpha^4}\bigg)\bigg(
60\mathbb{C}^4(\alpha-1)^3-60\mathbb{C}^2\pi^2(\alpha-1)^2(1+\alpha) \notag
\]
\[
\qquad +\pi^4(\alpha-3)(1+\alpha)(3+\alpha) +480\mathbb{C}(\alpha^3-1)\zeta (3)\bigg)
\bigg].  \notag
\]
But to  calculate the first four moments of $X^{'} $  we also need to know the moments of $W^{'}$.  The moments of $W^{'}$ can now be derived as follows:  The characteristic function of $W^{'}$ can be easily shown as
\[
\phi_{W^{'}}(t) = \mathbf{E} e^{itW^{'}}= \mathbf{E}W^{it}=\Gamma (\beta + i t)/ \Gamma (\beta),
\]
where $i=\sqrt{-1}.$ Using the logarithmic expansion of the gamma function (\cite{ans65}), we get the cumulant-generating function
\[
\log (\phi_{W^{'}} (t) )= \sum\limits_{k=1}^\infty \frac{(it)^k}{k!} c_k,
\]
where the $k$th cumulant is given by
\[
c_k=\psi^{(k-1)} (\beta), \quad \text{where} \quad  \psi^{(0)} (\beta) = \psi (\beta).
\]
Please note that the mean and variance of $W^{'}$ are given by
\[
\mu_{W^{'}}=c_1=\psi(\beta),\qquad  \text{and} \qquad \sigma_{W^{'}}^2=c_2= \psi^{(1)}(\beta),
\]
which are commonly known as the digamma and trigamma functions, respectively. For $k\geq3$, the $k$th cumulant  is the polygamma function of order $k-2$ evaluated  at $\beta$.  The  $k$th integer-order moments $\mathbf{E} (W^{'})^k$ can be calculated using the recursive relation
\[
\psi^{(k-1)} (\beta)=\mu_k^{'}- \sum\limits_{j=1}^{k-1} \binom{k-1}{j-1}c_j \mu_{k-j}^{'}.
\]
This implies that $\mu_1^{'} =\mu_U=c_1=\psi(\beta), \mu_2^{'} = c_2 +c_1^2=  \psi^{(1)}(\beta) + \psi(\beta) ^2 , \mu_3^{'}=c_3 +3c_2c_1 + c_1^3 = \psi^{(2)}(\beta) + 3\psi^{(1)}(\beta) \psi(\beta) + \psi(\beta)^3,  \mu_4^{'}=c_4 +4c_3c_1 +3 c_2^2 + 6c_2c_1^2 + c_4^3 = \psi^{(3)}(\beta) + 4 \psi^{(2)}(\beta) \psi(\beta) + 3 \psi^{(1)}(\beta)^2 + 6\psi^{(1)}(\beta)\psi(\beta)^2 + \psi(\beta)^4$, and so forth. We can now derive estimating equations using the first two moments of $X^{'}$. More specifically, it can easily be shown that the mean and variance of $X^{'}$ are
\begin{equation*}
\mu_{X^{'}}= \gamma\bigg( \frac{1}{\alpha} -1\bigg) + \frac{1}{\alpha}\psi (\beta), \tag{3.1}   \label{3.1}
\end{equation*}
and
\begin{equation*}
\sigma_{X^{'}}^2= \frac{\pi^2}{6}\bigg( \frac{1}{\alpha^2} -1\bigg) + \frac{1}{\alpha^2}\psi^{(1)} (\beta), \tag{3.2}  \label{3.2}
\end{equation*}
correspondingly. However, a more direct procedure is to consider  the characteristic function of the log-transformed random variable $X^{'}$. This simply suggests that
\[
\phi_{X^{'}}(t) = \mathbf{E} e^{itX^{'}}= \mathbf{E}e^{it(\frac{1}{\alpha} W^{'} + S_\alpha^{'}  )}= \mathbf{E}W^{it/\alpha} \mathbf{E}(S_\alpha)^{it}= \frac{\Gamma (\beta + i t/\alpha)}{ \Gamma (\beta)}\frac{\Gamma (it/\alpha)}{\alpha\Gamma (it)}.
\]
Taking the logarithmic expansion of the preceding equation yields the following cumulant-generating function of $X^{'}$
\[
\log (\phi_{X^{'}}(t))= \sum\limits_{k=1}^4 \frac{(it)^k}{k!} d_k + \sum\limits_{l=5}^\infty  \frac{(it)^l}{l!} d_l ,
\]
where the $k$th cumulant is given by
\[
d_k=\frac{1}{\alpha^k} \big[ \psi^{(k-1)} (\beta) + (-1)^k   \psi^{(k-1)} (1) (1- \alpha^k ) \big], \quad k=1,\ldots,4,  \; \text{and} \;  \psi^{(0)} (\tau) = \psi (\tau).
\]
It is easy to check that
\[
\mu_{X^{'}}=d_1=\frac{1}{\alpha} \big[ \psi(\beta) - \psi(1)(1- \alpha)\big], \;  \text{and} \; \sigma_{X^{'}}^2=d_2= \frac{1}{\alpha^2} \big[ \psi^{(1)}(\beta) + \psi^{(1)}(1) (1- \alpha^2)  \big],
\]
where $\psi (1) =  -\gamma$ and $\psi^{(1)}(1)=\pi^2/6$.  Also, using the recursive relation between the cumulant and the $k$th moment above, we can easily derive an expression for the third moment of the random variable $X^{'}$ as
\begin{align} \notag
\mathbf{E}(X^{'})^3 &= \frac{3\big[ (1/\alpha-1)^2\gamma^2 + \frac{\pi^2}{6}(1/a^2-1) \big] \psi(\beta) }{\alpha} +  \frac{3(1/\alpha-1)\gamma\big[ \psi(\beta)^2+ \psi^{(1)}(\beta)  \big]}{\alpha^2} \notag \\
&+ \frac{\psi (\beta)^3 +3\psi (\beta)\psi^{(1)}(\beta) + \psi^{(2)}(\beta)}{\alpha^3} \notag \\ \notag
&- \frac{2 (\alpha-1)^3  \gamma + \pi^2(\alpha-1)^2(1+\alpha)\gamma-4(\alpha^3-1)\zeta (3) }{2\alpha^3}.  \notag
\end{align}
The expression for the fourth moment easily follows  but we omit it here.  

%% file: s3b.tex
\section{Parameter estimation}

\subsection{Estimation for ML$(\alpha, \delta)$ distribution }

 One way of estimating the parameters of the ML distribution is to derive the method-of-moments estimators using formula (\ref{1.2}) for the fractional moments as in \cite{nas95} and \cite{koz01}. That is, select two values of $q < \alpha$,  $q_1$ and $q_2$ say, and compute $\hat \alpha$  and $\hat \delta$ numerically   using the following two equations:
\[
\hat{e}(q_l) = \frac{1}{n} \sum\limits_{i=1}^n  T^{q_l} = \frac{q_l \pi \hat{\delta}^{q_l}  }{ \hat{\alpha} \Gamma (1-q_l) \sin (\pi q_l/\hat{\alpha})}, \quad l=1,2.
\]
We re-emphasize that we need to choose appropriate numbers $q_1$ and $q_2$ beforehand, which are required to be less than $\alpha$  to be able to use the fractional moment estimators. This restriction suggests that we need to know  or have information about $\alpha$ a priori to be able to estimate the parameters $\alpha$ and $\delta$. 

To overcome this difficulty, we propose estimators of $\alpha$ and $\delta$ based on the mean and variance of the variable $T^{'}$. From (\ref{2.2}), the method-of-moments estimators for $\alpha$ and $\delta$ are
\begin{equation*}
\hat{\alpha}_P=\frac{2 \pi}{\sqrt{2(6 \hat{\sigma}_{T^{'}}^2 + \pi^2)}},\qquad  \text{and} \qquad \hat{\delta}_P =   \exp ( \hat{\mu}_{T^{'}} + \gamma), \tag{4.1}    \label{4.1}
\end{equation*}
respectively. Note that $\hat{\mu}_{T^{'}}$ and $\hat{\sigma}_{T^{'}}^2$ are the sample mean and variance of the log-transformed data $T^{'}$, correspondingly. Moreover, the preceding estimators are non-negative as desired. Another advantage of our estimation procedure is that it is computationally simple as we do not need to numerically find the unique solutions of a system of equations as the parameter estimates. The proposed estimators above are also shown to be asymptotically unbiased (see appendix).  

We now compare the proposed procedure with the fractional moment estimators ( $\hat{\alpha}_F$ and $\hat{\delta}_F$). In particular, we used  bias and root-mean-square error (RMSE) as bases for the comparison.  Following \cite{koz01}, we assumed $0.5 \leq \alpha < 1$, and used the same constants  $q_1=1/2 \leq \alpha$ and $q_2=1/4 \leq \alpha$.  The fractional moment estimator  $\hat{\alpha}_F$  of $\alpha$  is   computed by numerically solving the equation 
\[
\frac{\hat{e}\left(\frac{1}{2}\right) }{\left(\hat{e}\left(\frac{1}{4}\right)\right)^2} = \frac{ \left( \Gamma \left(3/4\right) \right)^2 8 \hat{\alpha}_F \sin^2 \left( \pi / 4 \hat{\alpha}_F \right) }{\pi^{3/2} \sin \left( \pi / 2 \hat{\alpha}_F \right)}.
\]
Then the fractional estimator of $\delta$ is  calculated as 
\[
\hat{\delta}_F = \frac{4 \hat{\alpha}_F^2 \sin^2  \left(\pi/2\hat{\alpha}_F\right) \frac{\left[\hat{e}\left(\frac{1}{2}\right)  \right]^2}{\pi} + \left[ 4 \hat{\alpha}_F \Gamma \left(3/4\right)   \sin  \left(\pi / 4\hat{\alpha}_F\right)  \frac{\hat{e}\left(\frac{1}{4}\right)}{\pi} \right]^4 }{2}.
\]
In the root calculation above, we used the \texttt{uniroot} function of the statistical software R with tolerance limit of $10^{-6}$.  We also performed 10000 simulation runs for each combination of the $\alpha$, $\delta$ and sample size $n$ values.   Table \ref{t1} in the appendix shows the computational test results. Clearly, the proposed estimators ($\hat{\alpha}_P$ and  $\hat{\delta}_P$) outperformed the fractional moment estimators ($\hat{\alpha}_F$ and  $\hat{\delta}_F$) even when the sample size is as large as 25000. When $n=25$, the bias ratio of the proposed $\hat{\alpha}_P$ to the fractional estimator $\hat{\alpha}_F$ ranges from 10.77\% to 48.64\%. This demonstrates the larger  bias  the fractional estimator is producing in estimating $\alpha$ for small samples. However, the bias difference seemingly becomes negligible as the sample size increases.  A similar observation  can be made for the bias incurred in estimating $\delta$. The RMSE's also generally shows that our procedure produces more homogeneous estimators that are closer to the true parameter values than the fractional moment method. These results certainly added another desirable characteristic of the proposed computationally simple approach.

\subsection{Estimation for GML$(\alpha, \beta)$ distribution}

We now propose  a similar estimation procedure for the  GML$(\alpha, \beta)$  distribution, and compare it with the fractional moment method.  Using the mean and variance of the log-transformed GML$(\alpha, \beta)$  distributed  random variable $X$ from Section 3, we can calculate parameter estimates  $\hat{\alpha}_P$ and $\hat{\beta}_P$ using the following two equations:
\begin{equation*}
\hat{\mu}_{X^{'}}= \gamma\bigg( \frac{1}{\hat{\alpha}} -1\bigg) + \frac{1}{\hat{\alpha}_P} \psi (\hat{\beta}_P), \tag{4.2}   \label{4.2}
\end{equation*}
and
\begin{equation*}
\hat{\sigma}_{X^{'}}^2= \frac{\pi^2}{6}\bigg( \frac{1}{\hat{\alpha}_P^2} -1\bigg) + \frac{1}{\hat{\alpha}_P^2} \psi^{(1)} (\hat{\beta}_P). \tag{4.3}  \label{4.3}
\end{equation*}
In this paper, we only consider an approximation  based on the first few terms of the series representation of the digamma function $\psi$ for performance comparison. A major advantage of using these estimating equations is that both digamma and trigamma functions are monotone in $\mathbb{R}^+$. Hence, finding the solutions is straightforward. Recall that
\[
\psi(\tau) = \log (\tau) -  1/(2\tau)  - 1/(12\tau^2) + 1/(120\tau^4)  - 1/(252\tau^6) + O(1/\tau^8).
\]
Thus, we approximate  $\psi(\tau)$ as 
\[
\hat{\psi} (\tau)= \log (\tau) -  1/(2\tau)  - 1/(12\tau^2) + 1/(120\tau^4)  - 1/(252\tau^6).
\]
This results to the following approximation of the trigamma function $\psi^{(1)} (\tau)$: 
\[
\hat{\psi}^{(1)} (\tau)= 1/ \tau  +  1/(2\tau^2) + 1/(6\tau^3) - 1/(30\tau^5) + 1/(42\tau^7).
\]
Solving the system of two equations  (\ref{4.2}) and (\ref{4.3}) using the preceding approximations of the digamma and trigamma functions will yield the parameter estimates  for the GML$(\alpha, \beta)$ distribution.

For the fractional moment technique,  we assumed $0.5 \leq \alpha <1$, $q_1=1/3$ and $q_2=1/4$  to compute $\hat{\alpha}_F$  and $\hat{\beta}_F$ numerically   using the two equations:
\[
\frac{1}{n} \sum\limits_{i=1}^n  X^{q_l} = \frac{ \Gamma \left( 1-q_l / \hat{\alpha}_F \right) \Gamma (\hat{\beta}_F + q_l/\hat{\alpha}_F)  }{ \Gamma (1-q_l) \Gamma (\hat{\beta}_F)}, \quad l=1,2.
\]
In the comparison, we used  the function \texttt{optim} in R for both procedures with identical tolerance limits ($10^{-6}$) and initial conditions. We also  generated 10000 random samples of size $n=25, 50, 100, 500, 25000$ each from the GML$(\alpha, \beta)$ distribution, and computed the bias and the root-mean-square error (RMSE).  The same conclusions from the preceding subsection  can be drawn from Table \ref{t2} in the appendix. The table clearly shows that the proposed procedure outperformed again the fractional moment method even for large sample sizes.  

Overall, Tables \ref{t1}-\ref{t2} in the appendix strongly indicate that  the proposed point estimators using the log-transformed data performed better in our comparisons.

\section{Concluding remarks}

We have derived  the first few moments of the log-transformed Mittag-Leffler distributed random variables. The log-moments led to systems of equations which are then used to estimate the parameters of the ML$(\alpha, \delta)$ and   GML$(\alpha, \beta)$  distributions.  A major advantage of our method over the other moment estimators (e.g., fractional moment estimators) is that we do not need to choose appropriate constants (e.g.,  $q_l<\alpha$) a priori to be able to calculate the parameter estimates. The calculations involved are straightforward. Approximate interval estimates for the parameters of the ML$(\alpha, \delta)$ distribution are derived. Furthermore, the testing and comparison have illustrated the superiority of our estimators.

Although some work have already been done, there are still a few things that need to be pursued. For instance, the development of estimators using Hill-type, regression, and likelihood approaches  would be a worthy pursuit. The application of these methods in  practice  particularly in economics and finance would be of interest also.

\newpage
\section{Appendix}

\subsection{Empirical results}

\begin{table}[h!t!b!p!]
\caption{\emph{Comparison of point estimators for the \text{ML}$(\alpha, \delta)$ distribution using different values of $\alpha$ and  $\delta$ for sample sizes $n=25, 50, 100, 500, 25000$.}}
 \begin{small}
 \centerline {
 \renewcommand{\arraystretch}{1.4}
\begin{tabular*}{6.45in}{|c@{\hspace{0.05in}}|c@{\hspace{0.05in}}|| c@{\hspace{0.05in}}| c@{\hspace{0.09in}}| c@{\hspace{0.05in}}|c@{\hspace{0.05in}} |c@{\hspace{0.05in}}||c@{\hspace{0.05in}}|c@{\hspace{0.05in}}|c@{\hspace{0.05in}}|c@{\hspace{0.05in}} |c@{\hspace{0.05in}}|}
 \hline
\multicolumn{2}{|c||}{}  &  \multicolumn{5}{|c||}{Bias}  &  \multicolumn{5}{|c|}{RMSE} \\ \hline \hline
$(\alpha, \delta)$ & $Est$ & $n=25$ & $50$ & $100$ & $500$ & $25000$ & $n=25$ & $50$ & $100$ & $500$ & $25000$\\ \hline \hline \hline
\multirow{4}{*}{$(0.5, 0.5)$} 
               &  $\hat{\alpha}_P$ & 0.018 & 0.010  & 0.004  & 0.001   & 0.000  & 0.085 & 0.059   & 0.041  & 0.018 & 0.002\\
               &  $\hat{\alpha}_F$ & 0.167 & 0.142 & 0.120   & 0.088   & 0.051  & 0.182 & 0.155   &  0.131  & 0.095   & 0.053\\ \cline{2-12}
               &  $\hat{\delta}_P$ & 0.135 & 0.061  & 0.029  & 0.007   & 0.000  & 0.519  & 0.296   & 0.188  & 0.078   & 0.010   \\
               &  $\hat{\delta}_F$ & 4.704 & 1.602  & 0.674  & 0.366   & 0.203  & 292.556 & 65.100   & 5.712  & 0.901   & 0.209 \\ \cline{2-12}               
\hline
\multirow{4}{*}{$(0.6, 5)$}  
               &  $\hat{\alpha}_P$    & 0.021   & 0.010  & 0.005   & 0.001   & 0.000  & 0.097 & 0.066   & 0.047  & 0.021   & 0.002   \\
               &  $\hat{\alpha}_F$    & 0.132   & 0.106  & 0.085   & 0.053   & 0.019  & 0.156  & 0.128   & 0.105  & 0.070   & 0.030  \\ \cline{2-12}
               &  $\hat{\delta}_P$    & 0.773   & 0.407  & 0.208   & 0.041   & 0.001  & 3.522  & 2.243   & 1.479  & 0.619   & 0.086   \\
               &  $\hat{\delta}_F$    & 6.915   & 2.562  & 1.731   & 1.079   & 0.395  & 334.817 & 22.053 & 4.053  & 4.043   & 0.589 \\ \cline{2-12}               
\hline
\multirow{4}{*}{$(0.7, 1)$}  
               &$\hat{\alpha}_P$   & 0.022   & 0.011  & 0.006   & 0.001   & 0.000  & 0.104 & 0.072   & 0.051  & 0.023   & 0.003   \\
               &$\hat{\alpha}_F$   & 0.099   & 0.076  & 0.058   & 0.031   & 0.006  & 0.131 & 0.107   & 0.089  & 0.059   & 0.024  \\ \cline{2-12}
               &$\hat{\delta}_P$   & 0.108   & 0.051  & 0.023   & 0.005   & 0.000  & 0.525 & 0.346   & 0.231  & 0.101   & 0.014   \\
               &$\hat{\delta}_F$   & 0.303   & 0.194  & 0.132   & 0.069   & 0.015  & 2.033 & 0.593   & 0.295  & 0.141   & 0.061 \\ \cline{2-12}               
\hline
\multirow{4}{*}{$(0.8, 100)$}  
               &$\hat{\alpha}_P$  & 0.020   & 0.010  & 0.005   & 0.001   & 0.000  & 0.105 & 0.075   & 0.053  & 0.024   & 0.003   \\
               &$\hat{\alpha}_F$  & 0.067   & 0.049  & 0.036   & 0.016   & 0.002   & 0.106 & 0.088   & 0.074  & 0.050   & 0.018  \\ \cline{2-12}
               &$\hat{\delta}_P$  & 6.957   & 3.547  & 1.808   & 0.343   & 0.001  & 41.680 & 28.209   & 19.214  & 8.405   & 1.189   \\
               &$\hat{\delta}_F$  & 13.208  & 8.406  & 5.572   & 2.226   & 0.283  & 45.167 & 29.343   & 19.732  & 9.572   & 3.128 \\ \cline{2-12}               
\hline
\multirow{4}{*}{$(0.9, 0.1)$}  
               &$\hat{\alpha}_P$  & 0.018   & 0.009  & 0.004   & 0.001   & 0.000  & 0.104 & 0.076   & 0.054  & 0.024   & 0.003   \\
               &$\hat{\alpha}_F$  & 0.037   & 0.027  & 0.018   & 0.007   & 0.000   & 0.078 & 0.066   & 0.056  & 0.038   & 0.013  \\ \cline{2-12}
               &$\hat{\delta}_P$  & 0.005   & 0.002  & 0.001   & 0.000   & 0.000  & 0.033 & 0.022   & 0.016  & 0.007   & 0.001   \\
               &$\hat{\delta}_F$  & 0.006   & 0.004  & 0.002   & 0.001   & 0.000  & 0.032 & 0.022   & 0.015  & 0.007   & 0.002 \\ \cline{2-12}               
\hline
\end{tabular*}
}
\end{small}
  \label{t1}
\end{table}

\begin{table}[h!t!b!p!]
\caption{\emph{Comparison of point estimators for the \text{GML}$(\alpha, \beta)$ distribution using different values of $\alpha$ and  $\beta$ for sample sizes $n=25, 50, 100, 500, 25000$.}}
 \begin{small}
 \centerline {
 \renewcommand{\arraystretch}{1.4}
\begin{tabular*}{6.6in}{|c@{\hspace{0.05in}}|c@{\hspace{0.05in}}|| c@{\hspace{0.05in}}| c@{\hspace{0.09in}}| c@{\hspace{0.05in}}|c@{\hspace{0.05in}} |c@{\hspace{0.05in}}||c@{\hspace{0.05in}}|c@{\hspace{0.05in}}|c@{\hspace{0.05in}}|c@{\hspace{0.05in}} |c@{\hspace{0.05in}}|}
 \hline
\multicolumn{2}{|c||}{}  &  \multicolumn{5}{|c||}{Bias}  &  \multicolumn{5}{|c|}{RMSE} \\ \hline \hline
$(\alpha, \beta)$ & $Est$ & $n=25$ & $50$ & $100$ & $500$ & $25000$ & $n=25$ & $50$ & $100$ & $500$ & $25000$\\ \hline \hline \hline
\multirow{4}{*}{$(0.5, 20)$} 
               &  $\hat{\alpha}_P$ & 0.019 & 0.009  & 0.005  & 0.001   & 0.000  & 0.083 & 0.059   & 0.042  & 0.019 & 0.003\\
               &  $\hat{\alpha}_F$ & 0.109 & 0.081  & 0.061   & 0.031   & 0.006  & 0.143 & 0.113   &  0.089  & 0.055   & 0.021\\ \cline{2-12}
               &  $\hat{\beta}_P$ & 4.955 &  2.382   & 1.220  & 0.243   & 0.004  & 13.013  & 7.942   & 5.221  & 2.0252   & 0.319   \\
               &  $\hat{\beta}_F$ & 34.751 & 23.6442  & 16.45  & 7.723   & 1.346  & 44.717 & 30.581   & 21.875  & 11.413   & 3.509 \\ \cline{2-12}               
\hline
\multirow{4}{*}{$(0.6, 15)$}  
               &  $\hat{\alpha}_P$    & 0.019   & 0.011  & 0.005   & 0.001   & 0.000  & 0.088 & 0.064   & 0.046  & 0.021   & 0.003   \\
               &  $\hat{\alpha}_F$    & 0.089   & 0.065  & 0.044   & 0.019   & 0.002  & 0.132  & 0.105   & 0.084  & 0.052   & 0.016  \\ \cline{2-12}
               &  $\hat{\beta}_P$    & 2.353   & 1.275  & 0.634   & 0.107   & -0.003  &  6.536  & 4.373   & 3.062  & 1.394   & 0.268   \\
               &  $\hat{\beta}_F$    & 12.363  & 8.548  & 5.850   & 2.459   & 0.235  & 16.378 & 11.895 & 8.900  & 4.849   & 1.462 \\ \cline{2-12}               
\hline
\multirow{4}{*}{$(0.7, 10)$}  
               &$\hat{\alpha}_P$   & 0.017   & 0.010  & 0.005   & 0.001   & 0.000  & 0.086 & 0.064   & 0.047  & 0.021   & 0.003   \\
               &$\hat{\alpha}_F$   & 0.069   & 0.048  & 0.033   & 0.012   & 0.000  & 0.117 & 0.096   & 0.078  & 0.048   & 0.013  \\ \cline{2-12}
               &$\hat{\beta}_P$   & 0.884   & 0.492  & 0.252   & 0.057   & 0.001  & 2.729 & 1.976   & 1.423  & 0.646   & 0.094   \\
               &$\hat{\beta}_F$   & 3.931   & 2.723  & 1.882   & 0.732   & 0.033  & 5.491 & 4.246   & 3.329  & 1.955   & 0.563 \\ \cline{2-12}               
\hline
\multirow{4}{*}{$(0.8, 5)$}  
               &$\hat{\alpha}_P$  & 0.013   & 0.007  & 0.003   & 0.001   & 0.000  & 0.079 & 0.058   & 0.043  & 0.020   & 0.002   \\
               &$\hat{\alpha}_F$  & 0.047   & 0.031  & 0.021   & 0.007   & 0.000   & 0.098 & 0.081   & 0.068  & 0.041   & 0.010  \\ \cline{2-12}
               &$\hat{\beta}_P$  & 0.177   & 0.091  & 0.049   & 0.012   & 0.000  & 0.817 &  0.576   & 0.422  & 0.192   & 0.028   \\
               &$\hat{\beta}_F$  & 0.730  &  0.497  & 0.337   & 0.127   & 0.002  &  1.262 & 0.974   & 0.801  & 0.502   & 0.132 \\ \cline{2-12}               
\hline
\multirow{4}{*}{$(0.9, 1)$}  
               &$\hat{\alpha}_P$  & 0.023   & 0.012  & 0.006   & 0.001   & 0.000  & 0.106 & 0.075   & 0.053  & 0.024   & 0.003   \\
               &$\hat{\alpha}_F$  & 0.031   & 0.019  & 0.012   & 0.003   & 0.000   & 0.085 & 0.067   & 0.054  & 0.031  & 0.006  \\ \cline{2-12}
               &$\hat{\beta}_P$  & 0.106   & 0.007  & 0.002   & 0.000   & 0.000  & 0.177 & 0.123   & 0.085  & 0.038   & 0.005   \\
               &$\hat{\beta}_F$  & 0.022   & 0.011  & 0.005   & 0.001   & 0.000  & 0.188 & 0.125   & 0.087  & 0.039   & 0.006 \\ \cline{2-12}               
\hline
\end{tabular*}
}
\end{small}
  \label{t2}
\end{table}

\newpage

\subsection{Interval estimation for the $ML(\alpha, \delta)$ distribution}

We first study the limiting distribution of our estimators $\hat{\alpha}$ and $\hat{\delta}$ for the $ML(\alpha, \delta)$ distribution. If we let
\[
\hat{\mu}_{T^{'}}= \overline{T^{'}} = \frac{\sum \limits_{j=1}^n T_j^{'}}{n} \quad
\text{and}\quad \hat{\sigma}_{T^{'}}^2= \frac{\sum
\limits_{j=1}^n \left(T_j^{'}-\overline{T^{'}}\right)^2}{n}
\]
then the following weak convergence holds \cite{fer96}, i.e.,
\[
\sqrt{n}\left(
  \begin{array}{c}
    \hat{\mu}_{T^{'}}-\mu_{T^{'}} \\
    \hat{\sigma}_{T^{'}}^2 - \sigma_{T^{'}}^2  \\
  \end{array}
\right) \stackrel{d}{\longrightarrow}  \textsl{\Large{N}} \left[
  \begin{array}{ccc}
    \left(
      \begin{array}{c}
        0 \\
        0 \\
      \end{array}
    \right)
   &, & \left(
       \begin{array}{cc}
         \sigma_{T^{'}}^2 & \mu_3^{'} \\
         \mu_3^{'} & \mu_4^{'}-\sigma_{T^{'}}^4 \\
       \end{array}
     \right)
    \\
  \end{array}
\right],
\]
as $n \to \infty$, where $\mu_3^{'}, \mu_4^{'},$ and $\sigma_{T^{'}}^2$ are defined in Section 2. Using a standard result on asymptotic theory, the weak convergence above implies that
\[
\sqrt{n}\left[\mathbf{g}(\hat{\bm{\theta}}_n)-\mathbf{g}(\bm{\theta})\right]\stackrel{d}{\to}N\left[\bm{0},\;\mathbf{\dot{g}}(\bm{\theta})^{'}\bm{\Sigma}\mathbf{\dot{g}}(\bm{\theta})\right],
\]
where  $\hat{\bm{\theta}}_n=(\hat{\mu}_{T^{'}},  \hat{\sigma}_{T^{'}}^2)^{'},  \mathbf{g}$ is a mapping from $\mathbb{R}^d \to\mathbb{R}^k$  and   $\dot{\bf{g}}(\bf{x})$ is continuous in a neighborhood of $\bm{\theta} \in \mathbb{R}^d$. We now apply this result to the consistent estimator of $\delta$.
Letting
\[
\mathbf{g}(\mu_{T^{'}},\sigma_{T^{'}}^2) = \exp \left( \mu_{T^{'}}+\gamma \right).
\]
The gradient then becomes
\[
\mathbf{\dot{g}}(\mu_{T^{'}},\sigma_{T^{'}}^2)= \left(
                                         \begin{array}{c}
                                            \exp \left( \mu_{T^{'}}+\gamma \right) \\
                                            0   \\
                                         \end{array}
                                        \right).
\]
This implies that
\[
\sqrt{n}\big( \hat{\delta}-\delta\big) \stackrel{d}{\longrightarrow}
\textsl{N} \left[0,\; \sigma_{\delta}^2 \right],
\]
where
\begin{align}
\sigma_{\delta}^2 &= \mathbf{\dot{g}}(\mu_{T^{'}},\sigma_{T^{'}}^2)^{'} \left(
       \begin{array}{cc}
         \sigma_{T^{'}}^2 & \mu_3^{'} \\
         \mu_3^{'} & \mu_4^{'}-\sigma_{T^{'}}^4 \\
       \end{array}
     \right)
     \dot{\bm{g}}(\mu_{T^{'}},\sigma_{T^{'}}^2) \notag \\
&=\frac{\pi^2 e^{ 2( \mu_{T^{'}} + \gamma)}}{6}\left( \frac{2}{\alpha^2}-1 \right)\\
&=\frac{\pi^2 \delta^2}{6}\left( \frac{2}{\alpha^2}-1 \right), \notag
\end{align}
where the last line is obtained by plugging in $( \log (\delta) - \gamma )$ for  $\mu_{T^{'}}$.  Similarly,
\begin{align}
\sqrt{n}\left(\hat{\alpha}-\alpha\right)&
\stackrel{d}{\longrightarrow} \textsl{N} \left[0, \;
\left(\frac{-3 \sqrt{2} \pi }{\left(6\sigma_{T^{'}}^2+\pi^2
\right)^{3/2}} \right)^2 \left(\mu_4^{'}-\sigma_{T^{'}}^4 \right) \right] \notag \\
& \stackrel{d}{\longrightarrow} \textsl{N} \left[0, \;
\frac{\pi^6\left( 32 -20\alpha^2 -\alpha^4\right)}{5\left(6\sigma_{T^{'}}^2+\pi^2 \right)^3\alpha^4} \right] \notag \\
& \stackrel{d}{\longrightarrow} \textsl{N} \left[0, \;
\frac{\alpha^2\left( 32-20\alpha^2-\alpha^4\right)}{40} \right],\notag
\end{align}
where the final simplification is attained by substituting $\sigma_{T{'}}^2=\frac{\pi^2}{6}\left( \frac{2}{\alpha^2} - 1\right)$,
\[
\mathbf{g}(\mu_{T^{'}},\sigma_{T^{'}}^2) = \frac{2 \pi}{\sqrt{2(6 \hat{\sigma}_{T^{'}}^2 + \pi^2)}},
\]
and
\[
\mathbf{\dot{g}}(\mu_{T^{'}},\sigma_{T^{'}}^2)= \left(
                                         \begin{array}{c}
                                           0 \\
                                           \frac{-3 \sqrt{2} \pi }{\left(6\sigma_{T^{'}}^2+\pi^2
\right)^{3/2}}    \\
                                         \end{array}
                                        \right).
\]
Therefore, we have shown that our method-of-moments estimators are
normally distributed (asymptotically unbiased) as the sample size $n$ goes large. Consequently, we can now
approximate the $(1-\varepsilon)100\%$ confidence interval for  $\alpha$ and $\delta$ as
\[
\hat{\alpha} \pm z_{\varepsilon/2}\sqrt{\frac{\hat{\alpha}^2\left(
32-20\hat{\alpha}^2-\hat{\alpha}^4\right)}{40 n}},
\]
and
\[
\hat{\delta} \pm
z_{\varepsilon/2}\sqrt{\frac{\pi^2 \hat{\delta}^2}{6 n}\left( \frac{2}{\hat{\alpha}^2}-1 \right)},
\]
respectively, where $z_{\varepsilon/2}$ is the $(1-\varepsilon / 2)$th quantile of the standard normal distribution, and $0 <\varepsilon<1$.